\documentclass[12pt]{article}
\usepackage{epsfig}

\usepackage{amssymb}
\usepackage{amsmath}
\usepackage{amsfonts}

\def\be{\begin{equation}}
\def\ee{\end{equation}}
\def\ba{\begin{array}{c}}
\def\ea{\end{array}}

\def\ben{$$}
\def\een{$$}

\newcommand{\bea}{\begin{eqnarray}}
\newcommand{\eea}{\end{eqnarray}}

\newcommand{\kt}{\rangle}
\newcommand{\br}{\langle}

\newcommand{\ed}{\end{document}}

\begin{document}

\title{Supersymmetric quantum  mechanics living on topologically nontrivial Riemann
surfaces}

\author{Miloslav Znojil\\
Nuclear Physics Institute ASCR,\\ 250 68 \v{R}e\v{z},\\ Czech
Republic\footnote{e-mail: znojil@ujf.cas.cz}
\\
.
\\
V\'{\i}t Jakubsk\'{y} \\
Departamento de F\'{\i}sica, \\
Universidad de Santiago de Chile,\\
 Casilla 307, Santiago 2,
Chile\footnote{e-mail: jakub@ujf.cas.cz } }

\maketitle

\newpage

\abstract{Supersymmetric quantum mechanics is constructed in a new
non-Hermitian representation. Firstly, the map between the partner
operators $H^{(\pm)}$ is chosen antilinear. Secondly, both these
components of a super-Hamiltonian ${\cal H}$ are defined along
certain topologically nontrivial complex curves $r^{(\pm)}(x)$
which spread over several Riemann sheets of the wave function. The
non-uniqueness of our choice of the map ${\cal T}$ between
``tobogganic" partner curves $r^{(+)}(x)$ and $r^{(-)}(x)$ is
emphasized. }

\vspace{2cm}

 KEYWORDS

 {supersymmetry, Schr\"{o}dinger equation, complexified coordinates}

\vspace{1cm}

 PACS

 {11.30.Pb, 03.65.Fd, 93.65.Db }

 \newpage

\section{Introduction and summary}

We intend to show how, on an overall background of quantum
mechanics, one could interconnect the purely algebraic concept of
supersymmetry (SUSY, cf., e.g., review \cite{Cooper}) with the
more or less purely analytic concept of quantum toboggans (cf.
refs.~\cite{tobo,alltobo,identifi,dva,Wesselse} or a compact
review paper \cite{toborev}).

The presentation of this material will be initiated by section
\ref{ddruha} on SUSY, followed by another introductory section
\ref{ttreti} on complexifications of coordinates in quantum
mechanics. On this background our main message will be delivered
in sections \ref{hlavni} and \ref{pohlavni}. We shall show how a
new class of SUSY representations can be constructed via models
where the ``coordinates" are defined as certain multisheeted
complex curves $r(s)$.

The mathematical core of our present message lies in a never
published observation that whenever one tries to intruduce
complexified partner coordinate curves $r(s)=r^{(\pm)}(s)$ in SUSY
context, a nontrivial ambiguity of this partnership arises during
the transition from the family of non-tobogganic curves (defined
inside a single Riemann sheet, i.e., in the mere cut complex
plane, $r(s)=r^{(0)}\in l\!\!\!C$) to the more general family of
the curves $r(s)$ which interconnect {\em several} Riemann sheets
of wave functions $\psi(r)$ in question.


\section{Supersymmetric quantum mechanics \label{ddruha} }

In review paper \cite{Cooper} one can find a virtually exhaustive
list of reasons for interest in the so called supersymmetric quantum
mechanics (SUSY QM). This list starts by the theory and
phenomenology of elementary particles where SUSY QM plays the role
of a methodical guide towards our understanding of its various
mathematical features \cite{Gelf,Witten,Wittenbe}. At the other end
of the list one finds a very close relationship of SUSY QM to
certain exactly solvable one-dimensional potentials $V_0(x)$ defined
most often as not too complicated functions of a single real
coordinate $x$.

From the purely mathematical point of view one can characterize
the majority of applications of SUSY QM  as  purely algebraic
constructions. One simply picks up two arbitrary linear operators
$A_0$ and $B_0$ acting in $I\!\!L^2(I\!\!R)$, say,
 \be
  A_0=
 -\frac{d}{dx} + W(x)\,,\ \ \ B_0=\frac{d}{dx} + W(x)
 \,
 \label{definition}
 \ee
where $W(x)$ is a not yet specified ``superpotential" and where
one introduces the following pair of the so called ``supercharges"
acting in $I\!\!L^2(I\!\!R)\bigoplus I\!\!L^2(I\!\!R)$,
 \be
{\cal Q}_0=\left [
 \begin{array}{cc}
0&0\\
B_0&0 \ea \right ], \ \ \ \ \ \ \tilde{\cal Q}_0=\left [
 \begin{array}{cc}
0& A_0
\\
0&0 \ea \right ]\,. \label{keyr}
 \ee
The set of the anticommutators of these two operators is then easily
shown to read
 \be
 \{ {\cal Q},\tilde{\cal Q}
\}={\cal H} , \ \ \ \ \ \ \{ {\cal Q},{\cal Q} \}= \{ \tilde{\cal
Q},\tilde{\cal Q} \}=0\,
 \label{labela}
 \ee
where we dropped subscripts and where we just have to add the
following definition of the so called supersymmetric Hamiltonian,
 \be
 {\cal H}={\cal H}_0=
 \left (
 \begin{array}{cc}
 A_0B_0&0\\
 0&B_0A_0
 \ea
 \right )\,.
 \label{jednamec}
 \ee
It is virtually trivial to verify the validity of the commutation
relation
 \be
[ {\cal H},{\cal Q} ]=[ {\cal H},\tilde{\cal Q} ]=0\,.
 \label{labelbe}
  \ee
New operators do not emerge so that the algebra is closed as one of
the simplest examples of supersymmetric algebra generated by the two
elementary (super)charges and single (super)Hamiltonian.

In the majority of the current applications of SUSY QM one usually
extracts the superpotential $W(x)$ from the most common
one-dimensional Schr\"{o}dinger equation
 \be
 -\frac{d^2}{dx^2}
 \,\Phi_0(x) +V_0(x)
 \,\Phi_0(x)=E_0
 \,\Phi_0(x)\,, \ \ \ \ \ \ x \in (-\infty,\infty)\,
 \label{gsse}
 \ee
for a ground-state wave function $\Phi_0(x)$ possessing no nodal
zeros. Eq.~(\ref{gsse}) is being re-read as an equivalent nonlinear
differential equation of the first order,
 \be
 V_0(x)-E_0=
 -\frac{d}{dx}
 W(x)+W^2(x)\,.
 \label{Ricati}
 \ee
This equation is satisfied by the superpotential $W(x)$ which can be
also defined by the following explicit formula,
 \be
 W(x)=-\frac{1}{\Phi_0(x)}\,\frac{d}{dx}\Phi_0(x)\,.
 \label{Ricfeati}
 \ee
Alternatively, with a given ground-state wave function $\Phi_0(x)$
the latter two equations can be perceived as an explicit definition
of the difference or, if you wish, of the zero-energy potential
$U_0(x) := V_0(x)-E_0$.

In ref. \cite{Cannata} we tried to analyze the consequences of a
transfer of the SUSY QM recipe (based on eqs.~(\ref{Ricati}) and
~(\ref{Ricfeati})) to the class of potentials
 \be
 \label{bugre}
 V(r) = V^{(BG)}(r)= -r^4 + {\cal O}(r^2)\,
 \ee
characterized by the ``asymptotically anomalous" behavior. In the
present paper we shall extend  the results of ref. \cite{Cannata}
to systems called quantum toboggans \cite{tobo,alltobo}.

\section{Quantum mechanics using complex coordinates $r$ \label{ttreti} }

In spite of an asymptotically repulsive character of the quartic
potential (\ref{bugre}) ``with wrong sign", Buslaev and Grecchi
proved the reality of the spectrum under certain assumptions
\cite{BG}. Moreover, they also explained why this spectrum remains
discrete and bounded below, i.e., in principle, observable.
Several further reasons for a thorough interest in the Buslaev's
and Grecchi's model $V^{(BG)}(r)$ may be found, e.g., in
ref.~\cite{Jones}.

In the language of physics one of the most unusual assumptions
connected with the use of eq. (\ref{bugre}) lies, incidentally, in
the manifest loss of the observability of the corresponding
``coordinate" $r$. Indeed, one is allowed to consider its various
complexified versions
   \be
  r = r_{\varepsilon}(x) = x - i\,\varepsilon(x)
   \label{bgsp}
  \ee
where $x \in (-\infty,\infty)$ is merely a parameter and where the
function $\varepsilon(x)$ has been chosen as an $x-$independent
positive constant in ref.~\cite{BG}). Other choices with
asymptotically growing $\varepsilon(x)$ were studied, e.g., in
\cite{Mateo}.

On this background another family of models called quantum
toboggans (QT, \cite{tobo}) may briefly be characterized by the
existence of a topologically nontrivial Riemann surface ${\cal S}$
on which one can define wave functions [say, $\Phi_0(x)$ of
eq.~(\ref{gsse})] by analytic continuation. This means that the
class of the non-QT curves $r(x)$ given by eq.~(\ref{bgsp}) is
being complemented by all the smooth QT curves $r^{(N)}(x)$ which
spread over {\em several}, $N >1$ sheets of ${\cal S}$.

One of the simplest illustrative examples of such a QT curve is
given by eq. Nr. (10) of ref.~\cite{identifi},
 \be
 r^{(N)}(x)=-{\rm i}\left [{\rm i}(x-{\rm i}\varepsilon)
 \right ]^{2N+1}\,,\ \ \ \ \
  x \in (-\infty,\infty)\,.
  \label{kurvicka}
 \ee
This formula describes a toboggan-reminding spiral which may only be
approximated by eq.~(\ref{bgsp}) at not too large $x$. Globally this
curve encircles the origin $N-$times. Thus, a nontrivial
$N-$dependence of the spectrum of energies may be expected to emerge
whenever one finds a branch point in $\Phi_0(r)$ at $r=0$.

In the same manner, the existence of several branch points in
$\Phi_0(r)$ would imply that one might consider a broader family
of curves $r^{(N)}_{(\varrho)}(x)$ where the symbol $\varrho$
should distinguish between their topologically non-equivalent
versions (a small demo is now available showing some
$r^{(N)}_{(\varrho)}(x)$ for two branch points \cite{Novotny}).

\section{Supersymmetric quantum mechanics using complex coordinates $r$
 \label{hlavni} }

The manifest impossibility of a return to the real-coordinate
limits $\varepsilon \to 0$ in eqs.~(\ref{bgsp}) or
(\ref{kurvicka}) will be of particular interest in SUSY QM. For a
typical asymptotic ground-state-like solution of Schr\"{o}dinger
equation with potential (\ref{bugre}) one would get the following
contradiction at $\varepsilon = 0$,
  \be
  \Phi_0(r)= \Phi_0^{(\pm)}(r)\ \sim \  \exp
 \left (\pm \frac{i\, r^3}{3} \right ) \ \notin \
 L_2(-\infty,\infty)\,, \  \ \ \ \
 \varepsilon = 0. \label{exannxx}
  \ee
Turning attention to $\varepsilon \neq 0$ we may follow the idea
presented in  ref.~\cite{Cannata} and start from one of the
following two initial or tentative choices of the wave function,
 \ben
\Phi_0(r)=\psi^{(-)}(x)= (x-i\varepsilon) \, \exp \left (
-i\,\frac{ (x-i\varepsilon)^3}{3} \right )\ \in \
L_2(-\infty,\infty),
 \een
 \ben
\Phi_0(r)=\psi^{(+)}(x)= \frac{1}{x+i\varepsilon} \, \exp \left (
+i\,\frac{ (x+i\varepsilon)^3}{3} \right )\ \in \
L_2(-\infty,\infty)\,.
 \een
In the light of eq.~(\ref{Ricfeati}) this postulate would lead to
the following two alternative, tentative superpotentials
 \be
W^{(\pm)}(x)=-{ \left [ \frac{d}{dx}\psi^{(\pm)}(x) \right ] /
\psi^{(\pm)}(x) } = \pm \left [
 \frac{1}{x\pm i\varepsilon}-
 i\,(x\pm i\varepsilon)^2 \right ]\,.  \label{susyp}
 \ee
Although the application of standard rules would lead to the
respective potentials
 \be
V^{(-)}(x)= -4i(x-i\varepsilon) -(x-i\varepsilon)^4, \label{pota}
 \ee
 \be
V^{(+)}(x)= \frac{2}{(x+i\varepsilon)^2} -(x+i\varepsilon)^4\,
\label{potbe}
 \ee
one can immediately verify that these potentials are {\em not}
related by the standard SUSY partnership~\cite{Cannata}. In the
latter reference it has been found that the above-constructed pair
of candidates for SUSY-related superpotentials satisfies a
modified relation
 \be
\left [W^{(+)} \right ]^2 - \left [\frac{d}{dx}W^{(+)} \right ] =
{\cal T}\left \{
 \left [W^{(-)} \right ]^2 + \left [\frac{d}{dx}W^{(-)} \right ] \right
\}{\cal T}\,. \label{modifi}
 \ee
The symbol ${\cal T}$ stands for an antilinear and involutive
operator of the usual complex conjugation. In the light of this
observation it has been found in~\cite{Cannata} that at a constant
and {\em positive} $\varepsilon \neq 0$ only the {\em
minus-superscripted} superpotentials are to be employed. Using
this knowledge one reveals that the standard SUSY QM formulism of
section \ref{ddruha} is still applicable, provided only that we
update definition (\ref{definition}) as follows,
 \be
  A=
 -{\cal T}\frac{d}{dx} + {\cal T}W^{(-)}(x)\,,\ \ \
 B=\frac{d}{dx}{\cal T} + W^{(-)}(x){\cal T}
 \,.
 \label{redefinition}
 \ee
Although these operators become antilinear, their use still leads to
the same algebraic consequences as above, giving not only the {\em
manifestly non-Hermitian} SUSY Hamiltonian
 \be
{\cal H}= \left [
 \begin{array}{cc}
H^{(-)}&0\\ 0&H^{(+)}\ea \right ]= \left [
 \begin{array}{cc}
  B^{} A^{}
&0\\ 0& A^{} B^{} \ea \right ]
 \ee
but also the corresponding supercharges in the current form
 \be
{\cal Q}=\left [
 \begin{array}{cc}
0&0\\
A^{}&0 \ea \right ], \ \ \ \ \ \ \tilde{\cal Q}=\left [
 \begin{array}{cc}
0& B^{}
\\
0&0 \ea \right ]\ . \label{keyb}
 \ee
This was the key observation made in ref.~\cite{Cannata}.
Unfortunately, its immediate transfer to QT models leads to
difficulties. Let us now show how one may get rid of them.

\section{SUSY in tobogganic models  \label{pohlavni} }

\subsection{An ambiguity of ${\cal T}$ in cut plane}

For toboggans the direct application of the method of section
\ref{hlavni} fails because the current use of the operator ${\cal
T}$ of complex conjugation requires that the coordinates remain
real,
 \be
 ({\cal T}\psi)(\vec{r}) := \psi^*(\vec{r})\,,
 \ \ \ \ \ \vec{r} \in I\!\!R^d\,.
 \ee
Then the antilinear operator ${\cal T}$ still can  be perceived as
interrelating the elements of the Hilbert space of states
$|\psi\kt$ with their dual partners $\br \psi |$.

Once we leave  the real space of $\vec{r}$ and, for the sake of
definiteness, once we turn attention just to the one-dimensional
example (\ref{bgsp}) we usually define our Hilbert space as a
space of quadratically integrable functions of variable $r$. In
such a case the operator of complex conjugation gets merely
slightly modified,
 \be
 ({\cal T}\psi)(r(x)) := \psi^*(r^*(x))\,.
 \ee
In this form, unfortunately, it moves, at least formally, our wave
functions out of the Hilbert space in which we started working.


\begin{figure}[h]
\begin{center}
\epsfig{file=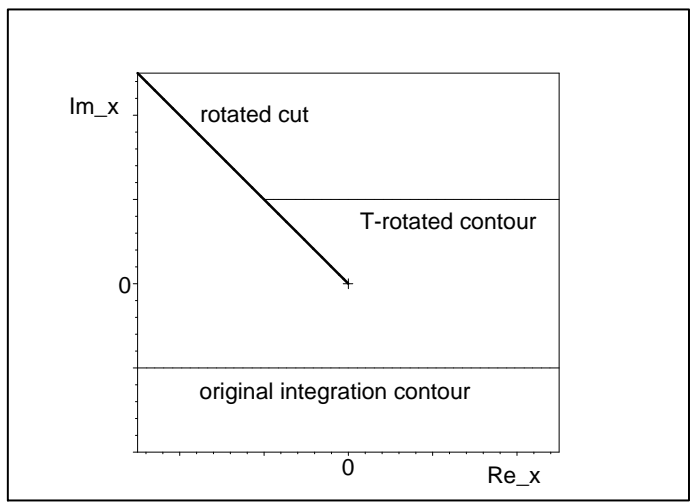,angle=270,width=0.6\textwidth}
\caption {The complex line of coordinates (\ref{bgsp}) and its
${\cal T}-$transformed version with ${\cal T}={\cal T}^{(+)}$. Our
upwards-oriented cut is slightly rotated to sample a part of the
next Riemann sheet.} \label{fione}
\end{center}
\end{figure}

The situation further worsens when your wave functions
$\psi(r(x))$ become tractable as analytic (i.e., in principle,
multivalued) functions of complex variable  $r\in l\!\!\!C$. In
Figure 1 we visualize the situation where the function $\psi(r)$
(defined along the complex contour (\ref{bgsp}) - see the lower
horizontal line in the picture) is assumed to possess a  branch
point located, say, in the origin of the complex $r-$plane. Then,
of course, we have to draw a cut from the origin (say, upwards)
and restrict our attention just to the resulting part of the
Riemann surface ${\cal S}$ of $\psi(r)$ (i.e., to its zeroth sheet
${\cal S}_0$). This {\em enables us} to ``see" all the curves
$r(x)$ \{as well as the functions $\psi[r(x)]$ which live on
them\}. At the same time, this {\em disables us} to ``see" all the
curve $r^*(x)$. In order to ``see" it \{and, of course, also the
${\cal T}-$image of our initial function $\psi[r(x)]$\} we would
have to move to one of the neighboring Riemann sheets (denoted,
conveniently, as ${\cal S}_{-1}$ and ${\cal S}_{+1}$).


\begin{figure}[h]
\begin{center}
\epsfig{file=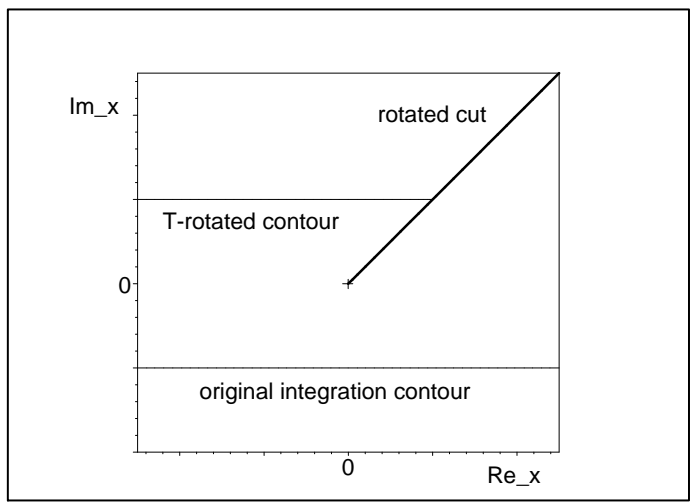,angle=270,width=0.6\textwidth}
\caption {The complex line of coordinates (same as in Figure
\ref{fione}) and its ${\cal T}-$transformed version with ${\cal
T}={\cal T}^{(-)}$. The cut is now slightly rotated in opposite
direction. } \label{fitwo}
\end{center}
\end{figure}

Schematically, these two possibilities are illustrated in Figures
1 and 2 where we see that in the presence of the branch point(s)
in $\psi(r)$, the ``complex conjugation" of the initial ``curve of
complex coordinates" $r(x)$ [exemplified by eq.~(\ref{bgsp})] can
be mediated by {\em at least two} nonequivalent formal operators
${\cal T}^{(\pm)}$ where the superscript indicates our choice
between the two eligible Riemann sheets ${\cal S}_{\pm 1}$ to
which we move. At the same time, it is also necessary to imagine
that for the generic, logarithmic form of ${\cal S}$  our
antilinear maps will also lose their involutive character so that
$({\cal T}^{(\pm)})^2 \neq I$ in general.

\subsection{The problem of classification of families of ${\cal T}$s
 on Riemann surfaces
 }

All the above considerations show that we must be very careful
with the notation conventions whenever a multivalued, analytic
wave function lives on some less trivial QT curve $\
r^{(N)}_{(\varrho)}(x)$. Once more we have to recollect the
dichotomy between the choice of the operators ${\cal T}={\cal
T}^{(+)}$ or ${\cal T}={\cal T}^{(-)}$ as encountered under
assumption that there exists just a {\em single} branch point in
${\cal S}={\cal S}^{[1]}$ (cf. our Figures 1 and 2).

One immediately imagines that for all the Riemann surfaces ${\cal
S}={\cal S}^{[M]}$ which are ``punctured" by more (i.e., in
general, $M \geq 1$) branch points the classification of all the
eligible QT paths $r(x)$ gets quickly very complicated. In
particular, the dichotomy illustrated by our Figures 1 and 2 must
be understood {\em locally} at $M >1$. The above-mentioned $M=1$
distinction between ${\cal T}={\cal T}^{(+)}$ and ${\cal T}={\cal
T}^{(-)}$ must be performed in the vicinity of {\em every} branch
point $r_{BP}^{[K]}$ with $K = 1,2,\ldots,M$. In this way a
superscripted $M-$component multiindex $\varrho=(\pm, \pm ,
\ldots, \pm)$ appears also in every specific choice of the
generalized complex conjugation,
 \be
  {\cal T}={\cal T}^{(\varrho)} \neq {\cal T}^{-1}\,.
  \label{genet}
  \ee
Marginally let us note that a set of nice analytic examples of
wave functions  where $M$ went up to five has been constructed by
Sinha and Roy \cite{SR} and that the amazing combinatorial
complications related to an exhaustive classification of the
non-equivalent QT paths already emerge in the first less trivial
case with $M=2$ \cite{dva}.

This being said, our final return to the QT SUSY QM is very easy
because once we overcome the combinatorial classification barriers
and choose any particular member ${\cal T}$ of the family of
conjugations (\ref{genet}) we only have to modify our
above-mentioned formula (\ref{redefinition}) and set
 \be
  A=
 -{\cal T}\frac{d}{dx} + {\cal T}W^{(-)}(x)\,,\ \ \
 B=\frac{d}{dx}{\cal T}^{-1} + W^{(-)}(x){\cal T}^{-1}
 \,.
 \label{reredefinition}
 \ee
In the purely algebraic language, all the rest of the construction
of the general QT-type SUSY generators remains fully analogous to
the above-described recipe. In contrast, in the language of
analytic functions the situation seems much more exciting due to
the flexibility of the resulting variability of the paths of
coordinates $r(x)$. At the same time, in is worth adding that any
more quantitative evaluation (say, of the related numerical
spectra of energies) still seems to be  a fairly difficult open
question at present~\cite{Wesselse}.

\section*{Acknowledgements}

MZ acknowledges Universidad de Santiago de Chile for hospitality
and GA\v{C}R grant Nr. 202/07/1307 and M\v{S}MT ``Doppler
Institute" project Nr. LC06002 for supplementary support. VJ was
supported by FONDECYT under grant  3085013. Collaboration also
partially supported by DICYT (USACH).

\end{document}